\documentclass[aip,graphicx]{revtex4-1}
\usepackage{graphicx}
\usepackage{subcaption}
\draft 

\begin{document}

The following article has been submitted to Review of Scientific Instruments. 
\bigskip

\title{An experimental study of inflow stability in a falling soap film} 



\author{Yuna Hattori}
\affiliation{Fluid Mechanics Unit, Okinawa Institute of Science and Technology}


\date{20 May 2024}

\begin{abstract}
Inflow control is essential for most fluid mechanics experiments. Although vertically falling soap film flows have been extensively used in the last four decades to study two-dimensional flows, its inflow stability has not yet been discussed in detail. In this article, aiming to improve the inflow stability of the system, we discuss how flow driving systems dominate the inflow states and the statistics of soap film flow properties. We report experimental measurements of inflow rates using different flow driving methods, followed by soap film velocity measurements by Laser Doppler Velocimetry (LDV). The widely-used method of a constant-pressure-head reservoir exhibits a continuous drop of inflow rate. In addition, even when the flow is not disturbed, the soap film displays high magnitudes of velocity fluctuations. The mean square of velocity fluctuation was measured at 4.6\% of the mean. We also test other methods without the reservoir, where the flow is directly controlled by a pump; two staggered syringe pumps are able to maintain inflow stability over days and reduce the velocity fluctuation to 0.5\% of the mean. From these measurements, we conclude that the drop of inflow rate might be caused by micro/milli-scale air bubbles, which cannot be completely removed in the system. A method to control the inflow actively is necessary in a soap film setup, when a stable and long standing film flow is needed.
\end{abstract}


\maketitle 


\section{\label{sec:intro}Introduction}
Controlling the inflow of a fluid mechanics experiment is crucial. Undesirable inflow instability or fluctuations can strongly affect the statistics of flow properties, such as the mean and root-mean-square velocities. Inflow stability is also known to have a significant role in the transition to turbulent state in pipe flow experiments~\cite{mullin2011experimental}. The biased data, in turn, perturbs our studies of the system.

In this article, we investigate the inflow of a vertically falling soap film setup. The soap film setup has been used extensively over the last four decades to study two-dimensional flows~\cite{soap_roughwalls,tran2010macroscopic,nature_soapfilm,cerbus2013intermittency,hidema2018effects,salkin2016generating,kellay2002two,tiwari2023effect,kim2021multiscale,korlimarla2020evolution}. The inflow rate usually ranges from 10 to 500 $\mu$l/s~\cite{rutgers2001conducting}. Such a small inflow rate is particularly sensitive and susceptible to small pressure changes.  The standard method to drive the soap film flow is to use a reservoir on top of the soap film, whose pressure head is set to be constant. However, with this method, we found several shortcomings: continuous drops of inflow rates, lack of reproducibility, and large mean square velocity fluctuations even in undisturbed flows. 

Our inflow measurements showed that the inflow rate is decreasing over time, which can be a few to 60 \% over 50 minutes. In fact, little attention has been paid to the inflow stability in previous soap film experiments. For example, a review paper on vertically falling soap film setups~\cite{rutgers2001conducting} covers a wide range of techniques, but the inflow stability over time was not investigated. One possible reason for the lack of inflow study is the limited duration of measurements; a typical measurement duration in soap film flows varies between a few seconds and a couple of minutes~\cite{soap_roughwalls,eshraghi2021flowing}. Another possible reason is that most of the previous studies investigated the films in turbulent states or other disturbed states, the inflow state was intertwined with the disturbance and therefore not observed independently. However, any inherent disturbance in the soap film flows might interact with the disturbance we create, potentially contaminating the flow dynamics we wish to study.

Repeatability of the same inflow rate is another problem. The inflow rate is typically controlled by a needle valve in the standard setup, but the same setting of the needle valve does not necessarily produce the same inflow rate every time. The reservoir is usually placed a short distance (e.g. 20 cm) above the injection nozzle, the flow driving pressure is therefore on the order of millibar (e.g. 20 mbar). Because of this, even a small change of the system, for example, a change in the number of air bubbles in the valve, can influence the inflow rate. In practice, this usually means the needle valve is adjusted right before each experiment to achieve the same inflow rate among experiments on different days or even the same day; this is neither rigorous nor repeatable. Lastly, we also found the mean square of velocity fluctuation of soap film flows was as large as 4.6\% of the mean, even when the flow is undisturbed. In general, to prevent laminar flows from transitioning to turbulence, one must reduce the amplitude of velocity fluctuations as much as possible. 

Here we present systematic inflow rate measurements in a vertically falling soap film flow driven by different methods; one is the standard method using a reservoir, and the others are by a pump without the reservoir. The advantage of connecting a pump directly is that we can control the inflow rates actively. Even when there are pressure drop changes, appropriate pumps can cover the pressure change. The pumps we tested are a diaphragm pump, a gear pump, and two staggered syringe pumps. The first two pumps produce disturbances themselves. The flows driven by two staggered syringe pumps showed the most stable and repeatable flows. For the inflow rate measurements, we used a thermal mass flow meter. To supplement those inflow rate measurements, we also conducted soap film velocity measurements by using Laser Doppler Velocimetry (LDV). 


\section{\label{sec:exp}Experimental setup}
\begin{figure}
\centering
    \includegraphics[scale = 1]{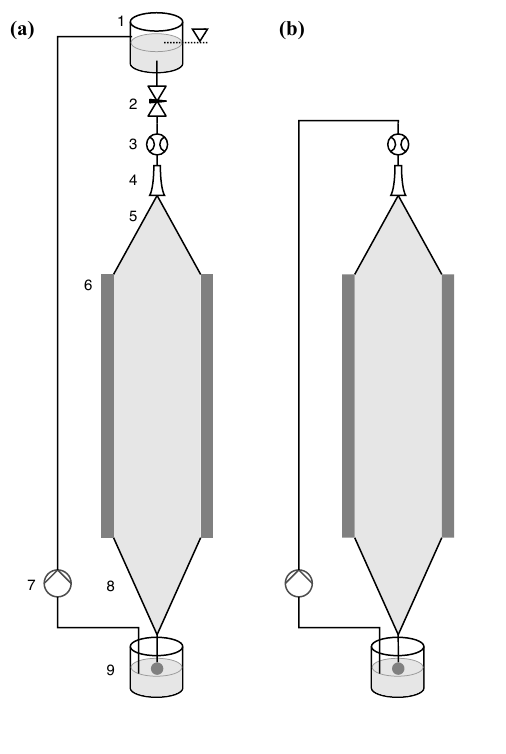}
    \caption{(a) The standard soap film setup with a reservoir. 1. top reservoir with overflow, 2. needle valve (Swagelok, SS-SS4-VH), 3. flow meter, 4. injection nozzle, 5. fishing wires, 6. 170 cm long steel blades in between which a soap film flow is made, 7. pump system that brings the soapy solution to the top reservoir, 8 and 9. fishing wires connected to a weight, to guide the film back to the bottom reservoir. PTFE tubes are used to connect each section (4 mm outer and 2 mm inner diameter). (b) Setup without a reservoir.}
    \label{fig:soapfilm_setup}
\end{figure}

The vertical soap film setup consists of three sections. The first section is a soap film flow section comprised of two parallel vertical blades (Fig.~\ref{fig:soapfilm_setup} (a), 6). The second section is an inlet and outlet section composed of wires that connect the blades to the feeding/collecting sections (Fig.~\ref{fig:soapfilm_setup} (a), 5 and 8). The third section is the soapy solution feeding/collecting section that is responsible for driving and maintaining a stable film flow (Fig.~\ref{fig:soapfilm_setup} (a) 1-4 and 7). This solution feeding section is the main focus in this article.

In the standard setup widely used by previous studies, the soapy solution feeding/driving section consists of a top reservoir and a pump (Fig.~\ref{fig:soapfilm_setup}, (a)). The level of the reservoir is kept constant by an overflow outlet at a specific height. In this study, after testing the standard setup, we removed the top reservoir and connect pumps directly to the nozzle (Fig.~\ref{fig:soapfilm_setup}, (b)). We discuss the stability for each method in detail in the next section~\ref{sec:flowstability}. The soapy solution is a 2\% (by volume) solution of a commercial dish-washing liquid soap (Dawn, ultra-concentrated antibacterial dish-washing liquid) mixed with water. Note that the soapy solution should be freshly made every time a measurement is started, otherwise the concentration of soap changes due to water evaporation. This might lead to a change of surface tension and viscosity in the soap film flow.

To monitor the flow stability, we used a Sensirion liquid flowmeter (SLQ-QT500). According to Sensirion, this sensor has the absolute accuracy of 5\%. This means the reported inflow rate is within 5\% of the actual inflow rate. The repeatability is 0.5\% of the reported inflow rate. We placed the flow meter just before the nozzle. In addition to the flow meter, we used Laser Doppler Velocimetry (LDV) to monitor the film velocities. We added 1$\:\mu m$ polystyrene particles (Sigma-Aldrich) to the soapy solution as tracking particles. The LDV equipment includes an Argon-ion laser (Coherent Innova 70C), a transmitter, frequency shifting optics, fiber manipulators, and a probe (Dantec FiberFlow system). 

\section{\label{sec:flowstability}Inflow stability}
\subsection{\label{subsec:topbucket}Reservoir system}
The reservoir system, with the outlet of the top reservoir connected to the nozzle with a needle valve in between (Fig.~\ref{fig:soapfilm_setup}(a)), is the most common system to drive the flow in vertical soap film experiments. How the reservoir is resupplied does not much affect the flow as long as the soapy solution is supplied at a higher inflow rate than is fed to the soap film. An overflow outlet is used to maintain a constant pressure head. The needle valve is used to adjust the inflow rate. Using the needle valve, we can achieve the small inflow rates from 10 to 500 $\mu$l/s required for soap film setups.

\begin{figure*}
\centering
    \includegraphics[scale = 0.8]{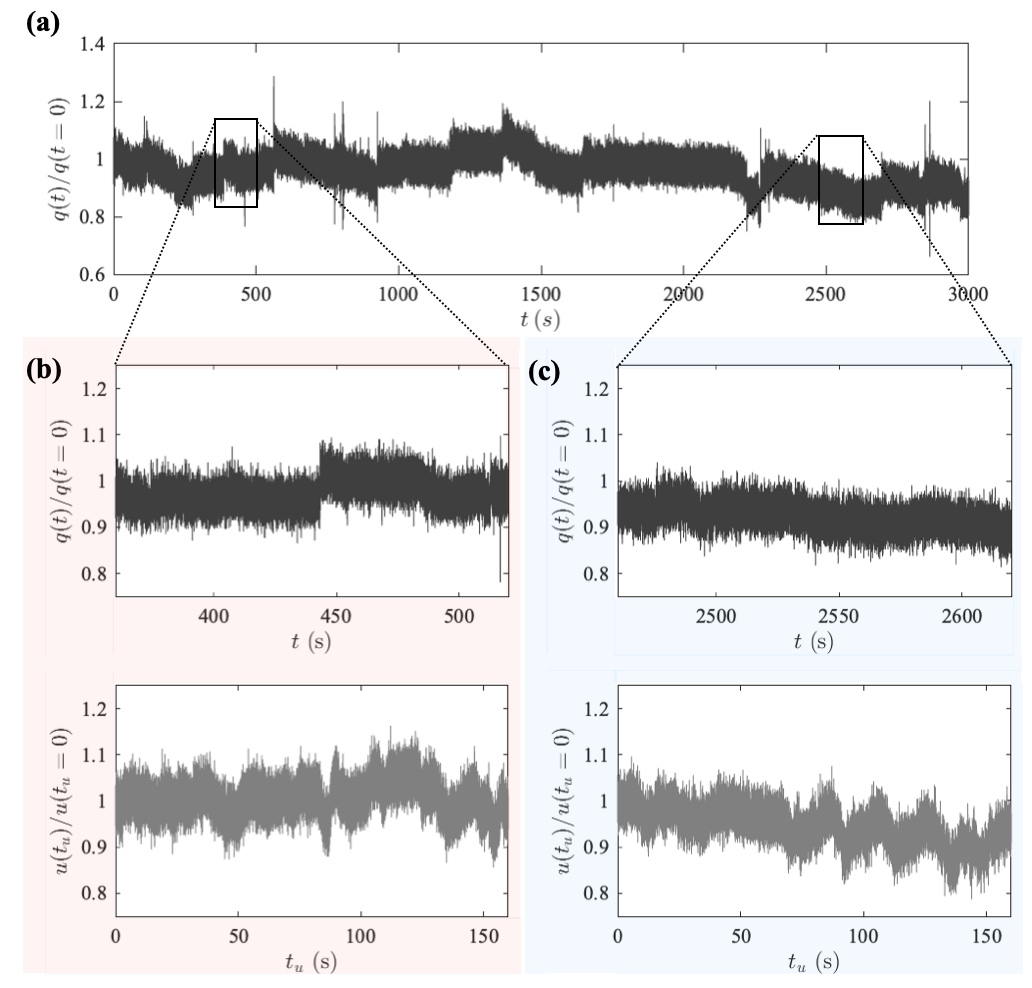}
    \caption{(a) The inflow rate of soapy solution normalised by the initial value, $q(t)/q(t = 0)$ vs. time $t$, with the top reservoir system. In this case, $q(t = 0) =$ 40 $\mu$l/s. The inflow fluctuates largely, and the mean value drops by 15\% over 3000 s. The data is taken long enough after starting the flow so the film has already reached a ``steady" state. (b) the top plot shows $q(t)/q(t = 0)$ vs. $t$, from 360 to 520 s. The bottom plot shows LDV measurement of soap film velocity, $u(t_u)/u(t_u = 0)$, taken simultaneously as the $q$ measurements. The LDV is placed at 18 cm downstream from the beginning of the blades along the center line of the film. The channel width of the soap film flow is 1.5 cm, $u(t_u = 0)$ is 1.52 m/s, and the rms is 4.6\% of the mean. (c) the top plot shows $q(t)/q(t = 0)$ vs. $t$, from 2460 to 2620 s. The bottom plot shows LDV measurement of soap film velocity, $u(t_u)/u(t_u = 0)$, taken simultaneously as the $q$ measurements.}
    \label{fig:topbucket}
\end{figure*}

\begin{figure}
    \centering
    \includegraphics{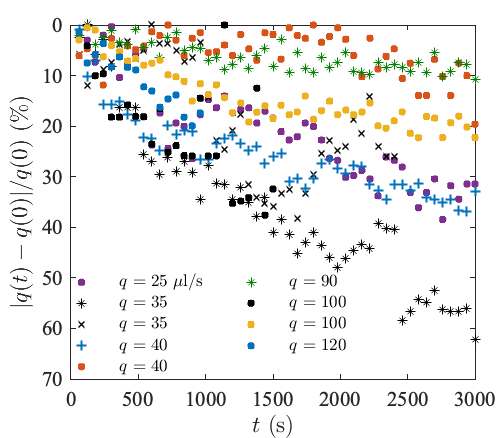}
    \caption{Drop of inflow rates described as $|q(t)-q(t = 0)|/q(0)$, vs. $t$, with the reservoir system. These nine different inflow rate measurements ($q$ in the legend is $q(t = 0)$) all show drops. }
    \label{fig:drop_stat}
\end{figure}

We conducted inflow rate $q(t)$ measurements with different magnitudes of inflow rates over a long period (Fig.~\ref{fig:topbucket} and Fig.~\ref{fig:drop_stat}). We found that the inflow fluctuates, and the overall inflow rate drops over time dramatically. The drops can range from a few percent to as high as 60\% over 50 minutes (Fig.~\ref{fig:drop_stat}). The inflow could be considered constant if we only observe a small time window, such as 200 s (Fig.~\ref{fig:topbucket}). However, for experiments that require stable flow on longer time scales, this flow driving method does not suffice. The soap film velocity $u(t)$ (Fig.~\ref{fig:topbucket}) taken simultaneously with the inflow rate measurement also showed the instability of the flow. Moreover, the root-mean-square of the velocity (Fig. 2b) was 4.6\% of the mean, although the flow is undisturbed.

\begin{figure}
\begin{subfigure}{0.45\textwidth}
    \caption{}
    \includegraphics[scale = 0.9]{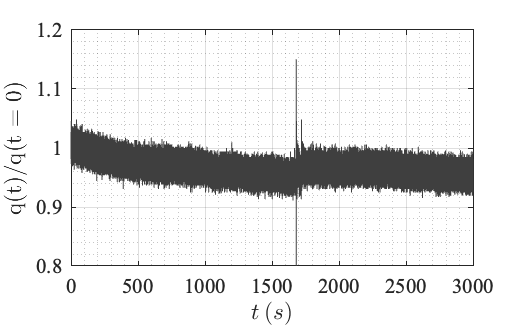}
    \label{fig:noParticle}
\end{subfigure}
\begin{subfigure}{0.45\textwidth}
    \caption{}
    \includegraphics[scale = 0.9]{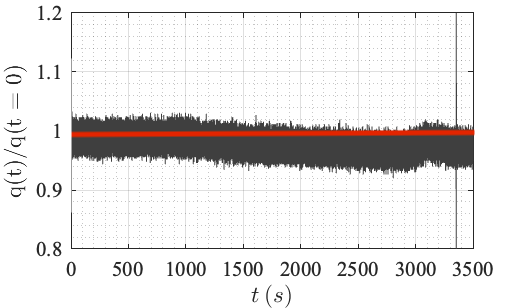}
    \label{fig:novalve}
\end{subfigure}
\begin{subfigure}{0.45\textwidth}
    \caption{}
    \includegraphics[scale = 0.9]{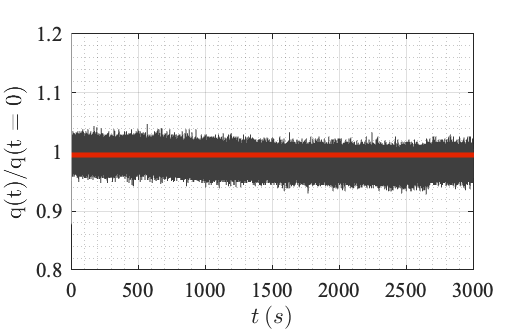}
    \label{fig:water}
\end{subfigure}
\caption{(a) $q(t)/q(t = 0)$ vs. time $t$, with the reservoir system, without seeding particles in the soapy solution ($q(t = 0) = 104\:\mu$l/s). The mean value still fluctuates, and drops 4\% over 3000 s. (b) $q(t)/q(t = 0)$ vs. time $t$, with the reservoir system, without a needle valve ($q(t = 0) = 103\:\mu$l/s). (c) $q(t)/q(t = 0)$ vs. time $t$, with the reservoir system using only pure water ($q(t = 0) = 114\:\mu$l/s).}
\end{figure}

This reservoir system seems to be the most reliable method, since it is driven by gravity. We may posit that the reason for the inflow rate drop is due to clogging by the tracer particles that are added for LDV measurements. To test this, we conducted measurements of inflow rates after thoroughly cleaning the setup without adding tracer particles in the soapy solution. However, the inflow rate still dropped (Fig.~\ref{fig:noParticle}). 

The next possible reason is needle valve. Needle valves for each inflow rates contain a very narrow flow restricting section (order of 1 $\mu$m) to produce the desired effect of flow control~\cite{xu1999experimental}. Not only seeding particles, but small dust particles, which the soap film might be collecting while flowing, might be narrowing or partially blocking the slit. We also observed air bubbles created after the needle valve, which could change the inflow rate. If the needle valve is the problem causing the inflow instability, we would not see it when we remove the needle valve. The inflow rate measurements without a needle valve is shown in Fig.~\ref{fig:novalve}. When removing the needle valve, we need to add a different flow restriction between the soap film and the nozzle. For this purpose we inserted 37 cm of a 1 mm inner diameter tube attached between the reservoir and the flow meter. This larger orifice flow restriction has the benefit of not clogging easily but obvious disadvantage that flow rate changes now require partially disassembly to adjust tube length. Small experiment to experiment adjustments become unfeasible. Although the inflow rate drop is much smaller, the inflow is still not constant over 3000 s. 

Another possibility is the influence of surfactants. As can be seen in Fig.~\ref{fig:water}, the drop of inflow rate is smaller when just using water. However, surprisingly, the continuous downward trend of the inflow rate persists. Note that tracer particles, needle valve, and surfactants are now removed, and the level of the reservoir is kept constant using the outerflow. 

Our hypothesis of why the continuous inflow rate drop happens is the following: water always contains some amount of dissolved gases. Some of this dissolved gas comes out of solution as the liquid undergoes pressure changes. Imagine tiny air bubbles are created and growing in the thin tubes of the soap film setup. According to the Hagen–Poiseuille equation, the inflow rate with a constant pressure gradient depends on the fourth power of the tube radius. This means, even when the size of air bubbles is only on the order of 100 $\mu$m, the change in the radius of the tube that the solution has to go through is non-negligible. Therefore, as the air bubbles grow, the inflow rate drops. Sometimes a bigger air bubble travels through the tubes and flushes out some of the tiny bubbles. This leads to a sudden increase of the inflow rate after continuous decrease of inflow rates (see Fig.~\ref{fig:water} around $t$ = 2700 s). Indeed, we observed more air bubbles at the end of the measurements. To solve this air bubbles issue, one might think of degassing the water before the measurements. However, in soap film measurements, the film is always interacting with the surrounding air, where gas can dissolve into the solution. Therefore, that is not a solution.

In conclusion, with the standard setup of a constant-pressure-head reservoir, the inflow is susceptible to many factors in the setup. Each factor is worth investigating in detail, but that is not the focus of the current article. These above investigations suggested us not to pursue the constant-pressure control, but to aim for direct control of inflow rates. The main focus in this article is having a stable inflow rate in a vertically falling soap film.

\subsection{Direct pump feeding system}
\begin{figure}
    \centering
        \includegraphics{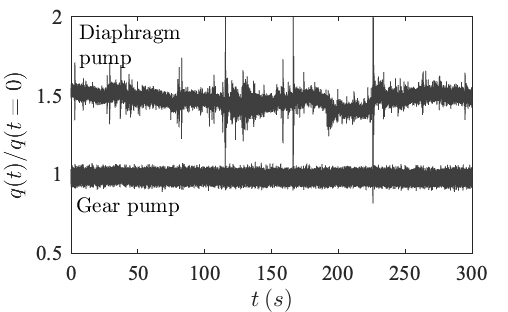}
    \caption{$q(t)/q(t = 0)$ vs. time $t$, with diaphragm pump ($q = 400\:\mu l/s$) and gear pump ($q = 500\:\mu l/s$). The time series for the diaphragm pump case is shifted by +0.5 to better show the time series in the same plot. The diaphragm pump produces a large fluctuation in the flow. The gear pump shows a stable mean but is pulsing by its nature (see Fig.~\ref{fig:spectra}). }
    \label{fig:gear_dia}
\end{figure}

In order to better control the inflow rate, now a pump is directly connected to the soap film, bypassing the top reservoir (Fig.~\ref{fig:soapfilm_setup}(b)). This way, we can actively control the inflow rate, instead of keeping the pressure head constant. To our best knowledge, there is only one study that used this direct pump feeding~\cite{liu2016janus}. In this study, a diaphragm pump was used and the flow rate was adjusted by pinching the tubes. Reproducing this method, which still lacks direct flow control, the inflow showed large fluctuations (Fig.~\ref{fig:gear_dia}). In addition, this way of controlling the inflow rate is again not easily reproducible. It is therefore important to choose an appropriate pump for this direct pump feeding system. 

The next pump tested was a gear pump, produced by ISMATEC (Fig.~\ref{fig:gear_dia}). The inflow rate produced by this pump is easily adjustable by changing the rotation speed. The time series of the flow rate visually appeared to be constant and not fluctuate. To scrutinize for any possible pulsations we cannot see in the inflow rate time series $q(t)$, we computed the power spectra of $q(t)$ by using the Welch periodogram. From this analysis, we can clearly see that the gear pump has an inherent pulsing, and the pulsing is more apparent with smaller inflow rates (see Fig.~\ref{fig:gear_spec}). At small inflow rates, the fluctuation was visually apparent even in the film. The peaks in the spectra of $q$ = 170 $\mu$l/s show the fundamental frequency of 4 Hz and its harmonics. This frequency lies well in the range of the rotation speed of the gear specified by ISMATEC at 50 to 500 rpm. Hence, this pump is also not suitable. The direct pump feeding systems thus far perform better than the reservoir system, but have their own drawbacks. We need a pump system that itself does not cause any disturbance to the flow, even at small inflow rates.

\begin{figure}
    \centering
    \includegraphics[width=8cm, height=10cm]{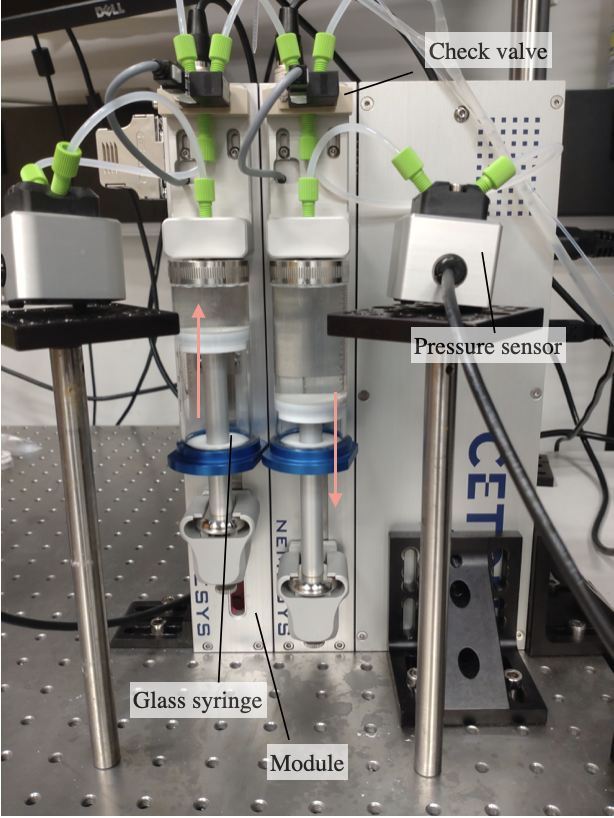}
    \caption{The CETONI neMESYS low-pressure syringe pump system with two modules, driving soapy solution to make a soap film flow. The system is set upright to release the air bubbles from the syringes. The pressure sensors are set closer to the syringe outlets and the check valves, to avoid air bubbles from being created in the soapy solution (when the soapy solution travel in a narrower tube, more air bubbles seem to be created. It is better to shorten the length of narrow tube region as much as possible).}
    \label{fig:cetoni}
\end{figure}

A common pumping system that is used in the field of microfluidics is a syringe pump. This option is highly attractive for soap film flows as they can provide a constant flow rate and pulsation-free flows even at small flow rates. To obtain a continuous, uninterrupted flow, we used the CETONI neMESYS low-pressure syringe pump system (Fig.~\ref{fig:cetoni}). Here, we utilized two modules which are both connected to a master controller. Each module consists of a high-quality glass syringe (50 ml, 8 bar; also produced by CETONI) driven by a translation stage, a 3x2 way electronically controlled valve, and a pressure transducer for monitoring the system pressure. If there were only one syringe, the experiment would be limited by the time it takes to drain the syringe. With two syringes, we can switch to the second syringe when the first syringe becomes empty. While the second syringe is feeding the film, we can fill up the first syringe, and keep the film continually fed. However, in this case, the switching of the syringes must be smooth. The two syringes switch smoothly between themselves by monitoring and matching the pressure values before switching (this function is called the pressure control continuous flow mode in the CETONI software, Qmix Elements). The pressure of the flow system (syringe + tubing close by) is kept at around 1.5 bar (it could change over a long run, due to air bubbles in the system as discussed in the Section.~\ref{subsec:topbucket}), which is the optimal pressure value CETONI recommends for having a smooth switch between syringes. To adjust the pressure drop towards the nozzle, we inserted different lengths of 1 mm inner diameter tubes in the loop. A needle valve is not suited to achieve this system pressure due to the aforementioned clogging problems with needle valves, also discussed in the Section.~\ref{subsec:topbucket}.

\begin{figure*}
    \centering
    \includegraphics[scale = 0.8]{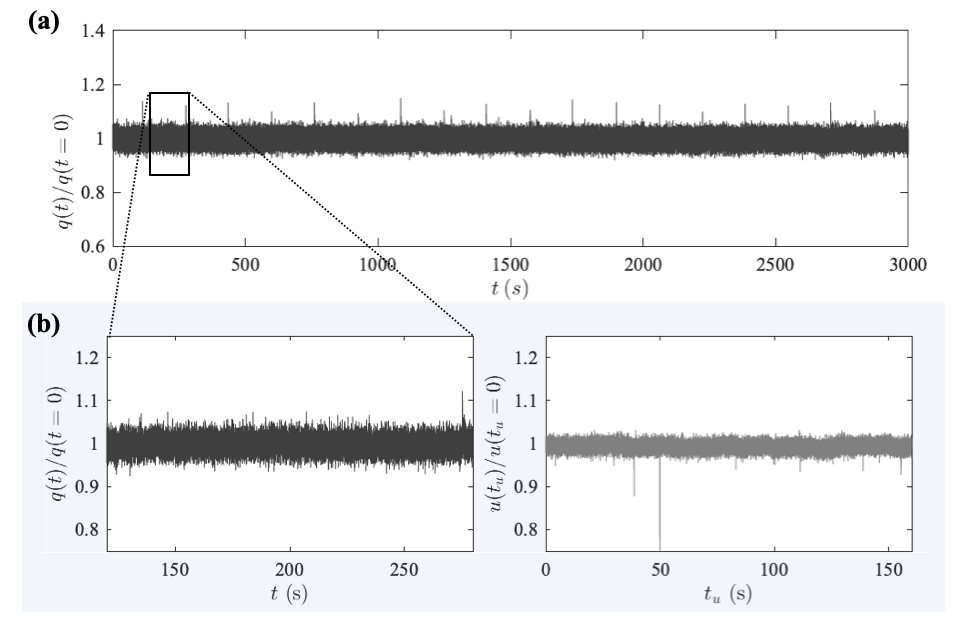}
    \caption{(a) $q(t)/q(t = 0)$ vs. time $t$, with the syringe pump system. In this case, $q(t = 0)$ is 85 $\mu$l/s. The inflow rate is extremely constant semi-infinitely. (b) the left shows $q(t)/q(t = 0)$ vs. time $t$, from 120 to 280 s. The right shows LDV measurement of soap film velocity, $u(t_u)/u(t_u = 0)$, taken simultaneously as the $q$ measurements. The LDV is placed at 30 cm downstream from the beginning of the blades and along the center line of the film. The channel width of the soap film flow w is 1.5 cm. In this case, $u(t_u = 0)$ is 2.50 m/s. The magnitude of rms is 0.5\% of the mean. }
    \label{fig:syringe}
\end{figure*}

\begin{figure}
    \centering
    \includegraphics[scale=1]{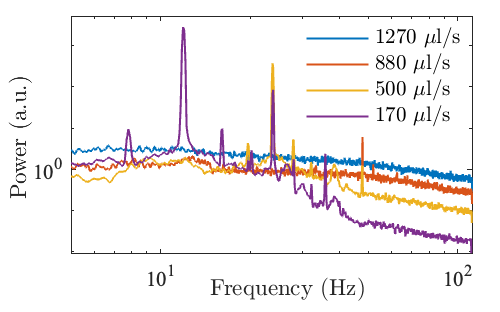}
    \caption{Power spectra of inflow rates, $q(t)$, with the gear pump. Each line shows a different inflow rate. }
    \label{fig:gear_spec}
\end{figure}

\begin{figure}
    \centering
    \includegraphics{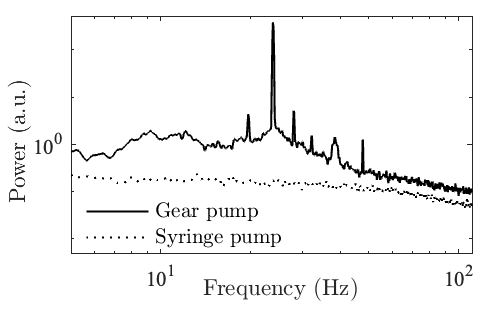}
    \caption{Power spectra of inflow rates, $q(t)$. The solid line is produced with the gear pump ($q = 500\:\mu l/s$), and the dotted line with the syringe pumps ($q = 500\:\mu l/s$). The gear pump clearly produces a large pulsing in the flow. On the other hand, the syringe pumps produce no pulsing.}
    \label{fig:spectra}
\end{figure}

The inflow rate with this method, and soap film velocity are shown in the Fig.~\ref{fig:syringe}. Both are extremely stable as long as the syringe pumps are moving. Note that the rms of velocity using this system is 0.5\% of the mean, which is significantly smaller than the one with reservoir system (4.6\% as mentioned in Section.~\ref{subsec:topbucket}). The spectrum also shows no relevant pulsation (Fig.~\ref{fig:spectra}). Thus, we have found a pumping system suitable for laminar soap film flow experiments.

Although air bubbles do not influence the inflow rate with this method, one should always try to minimise the amount of bubbles. This is because the bubbles could disturb LDV measurements in the soap film. If the plunger of a syringe is retracted too quickly, more air bubbles are produced as dissolved gas comes out of solution. With this limitation to the plunger retraction speed, we were able to reach up to 500 $\mu$l/s with the two syringe systems. If we stack more syringes, this flow rate can reach to higher values. Most importantly, the inflow rate is constant even with the existence of air bubbles in this setup, since we are controlling the inflow rate directly and actively instead of only setting the pressure head.

\section{\label{sec:conclusion}Conclusion}
The vertically falling soap film setup is simple, however, it requires careful attention to details to have a stable soap film flow over a long time. This article showed that the standard setup of a falling soap film driven by a reservoir system is stable only for a short period of time, such as the order of minutes. The inflow rates, in general, dropped continuously over long runs. The root mean square of soap film velocity was over 4\% of the mean inflow, which is too high for delicate laminar flow experiments. In fact, little attention has been paid to the inflow stability in the falling soap film. We conducted systematic measurements of the inflow rates and soap film velocities by using a thermal mass flow meter and Laser Doppler Velocimetry (LDV). Both results showed the instability of a falling soap film when the film flow was driven by the reservoir system. We also investigated and concluded that the inflow rates are susceptible to many factors such as micro/milli-scale air bubbles. When the inflow rate and pressure were actively controlled by a reliable pumping system such as staggered syringe pumps, both inflow rates and soap film velocities were stable over many hours. Importantly, with this active flow driving method, the rms of the soap film velocity was reduced to 0.5\% of the mean. Our conclusion is that a vertically falling soap film setup demands an active control of inflow rates, if a stable film flow for a long time or a very repeatable film flow is required.

\section*{Acknowledgments}
This study was supported by Okinawa Institute of Science and Technology Graduate University (OIST). I would like to acknowledge my Ph.D. supervisor, Prof. Pinaki Chakraborty and my colleagues, Dr. Julio M. Barros Jr., Mr. Christian Butcher, and Mr. Florian Fritsch for their immense help and fruitful discussions.

\section*{Author declarations}
\subsection*{Conflict of Interest}
The authors have no conflicts to disclose.

\section*{data availability}
The data that support the findings of this study are available from the author upon reasonable request.
\\

%
%

%


\bibliography{main}

\end{document}